\documentclass[aps,pre,twocolumn,showpacs]{revtex4}
\usepackage{graphicx}

\begin{document}

% Macros
\newcommand{\ub}{{\mathcal U}_{\rm bend}}
\newcommand{\us}{{\mathcal U}_{\rm stretch}}
\newcommand{\h}{\langle h\rangle}
\newcommand{\tn}{\theta_{0}}

\title{Modeling the elastic deformation of polymer crusts formed by sessile droplet evaporation}

\author{D. A. Head}	% and ...
\affiliation{Department of Applied Physics, Faculty of Engineering, University of Tokyo,
7-3-1 Hongo, Bunkyo-ku, Tokyo 113, Japan}

\date{\today}

\begin{abstract}
Evaporating droplets of polymer or colloid solution may produce a glassy crust at the liquid--vapour interface, which subsequently deforms as an elastic shell. For sessile droplets, the known radial outward flow of solvent is expected to generate crusts that are thicker near the pinned contact line than the apex. Here we investigate, by non--linear quasi--static simulation and scaling analysis, the deformation mode and stability properties of elastic caps with a non--uniform thickness profile. By suitably scaling the mean thickness and the contact angle between crust and substrate, we find data collapse onto a master curve for both buckling pressure and deformation mode, thus allowing us to predict when the deformed shape is a dimple, mexican hat, and so on. This master curve is parameterised by a dimensionless measure of the non--uniformity of the shell. We also speculate on how overlapping time-scales for gelation and deformation may alter our findings.
\end{abstract}


\maketitle

%
% INTRODUCTION
%
\section{Introduction}
\label{s:introduction}

The emergence of cost--effective microfluidic devices allowing manipulation and control of fluids on micrometer scales has promised a significant new paradigm for the manufacturing industry~\cite{Cuk:2000,Gans:2004}. Known as {\em bottom--up} processing, the required material can be transported to its destination in solvent form and deposited onto the target substrate using established technologies such as ink--jet head fluid ejection. One constraint of this means of material transport is that the deposited solution will evaporate, inducing non--trivial solvent flow that may distribute the solute in an irregular and undesirable manner. Understanding the physical mechanisms underlying such phenomena are thus crucial to obtaining the desired deposit profile.

Of concern here is the formation of a glassy `crust' that can appear on the liquid--vapour interface of evaporation polymer~\cite{Pauchard:2003a,Pauchard:2003b,Gorand:2004,Kajiya:pre,Pauchard:2003c} or colloid~\cite{Pauchard:2003c,Pauchard:2004,Tsapis:2005} solutions. Such crusts are believed to produce roughened surface profiles after drying of thin films~\cite{deGennes:2002}. For sessile  ({\em i.e.} pinned contact line) droplets on a wetting substrate, the effect is compounded by the outward radial flow of solvent during evaporation which, for low--concentration solutions, leads to enhanced solute deposition near the pinned line~\cite{Kajiya:pre,Picknett:1977,Deegan:1997,Hu:2002,Ozawa:2005}. For the higher concentrations where crust formation occurs, this outward flow suggests that the thickness of the crust will vary non--uniformly over the surface of the droplet, being thicker near the outer contact line and thinner near the centre or apex.

Being elastic, such a crust will deform under the internal osmotic pressure generated by continued solvent evaporation through its porous surface, and thus the final deposited polymer profile will at least partly depend on the mechanical properties of the crust. Assuming it to be everywhere thin, the relevant area of elasticity required is shell theory, an established field with known results for many different geometries, including the spherical caps relevant here (the droplet surface is spherical, so the initial crust profile should be approximately so); see {\em e.g.} Shilkrut~\cite{ShilkrutBook} for a recent survey. However, most of the literature is for shells of uniform thickness~\cite{Uddin:1987,KaiYuan:1994a,KaiYuan:1994b,Grigolyuk:2003} or specific, engineered non--uniformities~\cite{Simitses:1975}; those with a thickness profile similar to that expected for evaporating droplets have not been investigated. Furthermore, the rich array of deformed shell profiles observed in experiments~\cite{Pauchard:2003a,Pauchard:2003b,Gorand:2004,Kajiya:pre,Pauchard:2003c} has not, to the best of our knowledge, been systematically quantified as a function of the shell parameters, even for uniform shells.

\begin{figure}
\includegraphics[width=8.5cm]{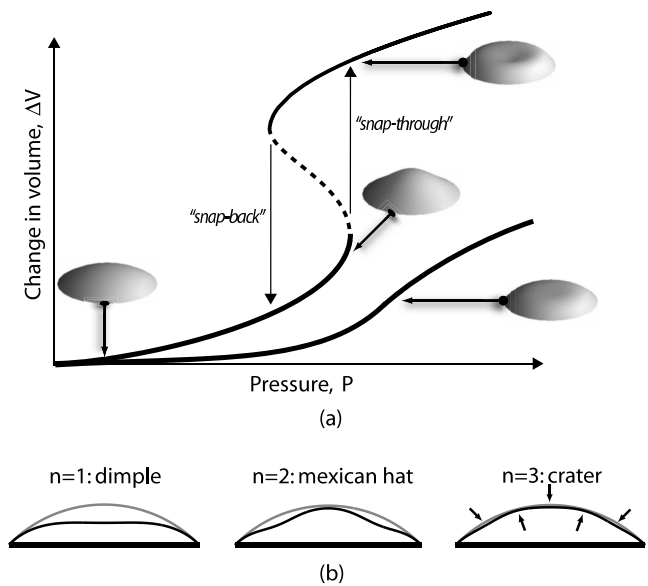}
\caption{
{\em (a)} Schematic equilibrium $P-\Delta V$ curves, one that buckles (with a dashed unstable regime), showing a hysteresis curve for cyclically--varying~$P$, and one that does not. Sketches of possible shell shapes are also shown. {\em (b)}~Lateral cross sections of pre--buckled shells (black) compared to the undeformed shape (grey), showing the first 3 deformation modes. For the crater mode, the turning points of the profile are indicated by arrows.}
\label{f:schematic}
\end{figure}

In this paper we describe numerical and theoretical investigations into the deformation of closed, elastic spherical caps with a thickness profile expected of crust formation during droplet evaporation, namely thin near the apex and thicker near the contact line. We restrict ourselves here to axisymmetric deformations that preserve the axis of symmetry of the shell; asymmetric deformations, as sometimes seen in experiments~\cite{Pauchard:2003a}, will be the subject of a future study. An overview of the problem is given in Fig.~\ref{f:schematic}(a). Here, two schematic equilibrium curves of inward pressure $P$ and change in droplet volume $\Delta V$ are given, one monotonic and one non--monotonic (here and below we define $P$ and $\Delta V$ to be positive for the deformations of interest). The S--shaped non--monotonic curve exhibits a buckling instability at a value~$P_{\rm c}$, when the shell jumps to an approximately inverted shape with a boundary layer. This is known as {\em snap--through}, or simply {\em snap} buckling, and can also be realised by {\em e.g.} applying a localised load~\cite{ShilkrutBook} or long--range attractive force~\cite{Tamura:2004}. If the pressure is subsequently decreased, a lower critical pressure is reached when the shell jumps to the original solution branch. This instability is known as {\em snap--back} buckling. The two differing critical pressures gives a hysteresis curve, the integrated area of which corresponds to the combined energy dissipated during both dynamic buckling events as the shell is damped to the static, equilibrium curve.

The above discussion assumes the pressure $P$ is being controlled. In fact, for the evaporation problem it is more natural to consider a slowly varying~$\Delta V$, which may exhibit different limit points; indeed, the simple schematic curves given in the figure are always stable under controlled volume. Nonetheless we shall here control~$P$, and focus much of our attention on the snap--through buckling point. The principle reason for doing this is that we are most interested in the deformation of the shell once non--linear effects have first become established, and the snap buckling point is a convenient point to focus on. Indeed, as shown below, the mode of the shell remains essentially unaltered from when non--linear effects first become significant to the buckling point itself, so we are effectively probing a broad range of the $P-\Delta V$ curve. Furthermore, controlling $P$ permits a closer correspondence with the spherical cap literature, for which a controlled volume is not typically considered. It also avoids the need for more sophisticated (and time--consuming) incremental--iterative algorithms~\cite{Yang:2003}. %It should be stressed that the {\em same} static $P-\Delta V$ curve is probed regardless of the choice of control parameter.

This paper is arranged as follows. In section~\ref{s:model} the model and simulation method are described. Results are then presented in section~\ref{s:results}, where it is shown that data for $P_{\rm c}$ as a function of the undeformed shell geometry can be made to collapse onto a single curve when scaled accordingly. The same scaled parameters also determine the shape of the shell, as quantified by the `mode number' $n$ (see Fig.~\ref{f:schematic}(b) for some examples). By parameterising shells of non--uniform thickness in a convenient way, we show that the effect of non--uniformity is to distort these master curves, without altering the gross underlying behaviour. This is supported by the scaling analysis of section~\ref{s:scaling}. Finally, we return to the droplet evaporation problem in light of our findings in section~\ref{s:discussion}.

%
% DEFINITION OF MODEL
%
\section{Model and simulation method}
\label{s:model}

To focus on the behaviour of elastic shells of non--uniform thickness, we shall here reduce the complex process of crust formation into a simple, idealised form. In particular, modulation of the liquid--vapour interface during crust formation, crust thickening after its initial genesis, and the possibility of temporally--evolving intrinsic curvature are all ignored. In essence we assume that the all of the polymer rapidly gels at the surface of the droplet, and is initially in a stress--free state. Since liquid droplets of size less than a few $mm$ have spherical surfaces~\cite{CapillarityBook}, our initial, undeformed shell configuration is taken to be a spherical cap with radius of curvature $R$ and a contact angle (the angle the shell makes with the substrate) $\theta_{0}$. This curved undeformed state identifies the crust as a {\em shell}, distinct from a plate which is flat when not deformed~\cite{Landau&Lifshitz}.

The static configuration of a thin shell can be purely defined in terms of the geometry of the mid-plane surface ${\mathcal S}$ and a (possibly variable) thickness $h$, defined such that the outer and inner surfaces lie at distances $\pm h/2$ normal to~${\mathcal S}$.  As we are only concerned here with quasi--static deformations, it suffices to find shell configurations that minimise ${\mathcal U}-P\Delta V$, with ${\mathcal U}$ the elastic strain energy and $-P\Delta V$ is the work done by the pressure~$P$ (in our sign notation). It is convenient to partition ${\mathcal U}$ into two parts, a stretch term

\begin{equation}
U_{\rm stretch}=\frac{1}{2}\int{\rm d}{\mathcal S}\,\frac{Eh}{1-\nu^{2}}\left\{
u_{\theta\theta}^{2}+u_{\phi\phi}^{2}+2\nu u_{\theta\theta}u_{\phi\phi}
\right\}
\label{e:U_stretch}
\end{equation}

\noindent{}where $u_{ij}$ is the strain of the mid-plane, in spherical coordinates in which $\theta\in[0,\theta_{0}]$ is the azimuthal and $\phi\in[0,2\pi]$ the polar angles, resp., $E$ is the Young's modulus of the material, and $\nu$ its Poisson ratio. In addition, there is a bending energy term

\begin{equation}
U_{\rm bend}=\frac{1}{2}\int{\rm d}{\mathcal S}\,\frac{Eh^{3}}{12(1-\nu^{2})}\left\{
\kappa_{\theta\theta}^{2}+\kappa_{\phi\phi}^{2}
+2\nu \kappa_{\theta\theta}\kappa_{\phi\phi}
\right\}
\label{e:U_bend}
\end{equation}

\noindent{}where $\kappa_{ij}$ the {\em change} in curvature of the shell meridian from the undeformed value $1/R$~\cite{WashizuBook,SeideBook}.  Terms proportional to $u_{\theta\phi}$ or $\kappa_{\theta\phi}$ vanish for axisymmetric deformations and have been dropped.

%$P$ and $p$ are defined as positive for positive external pressure or negative internal pressure, and so will always be positive for the osmotic driving of interest here. 

The expressions (\ref{e:U_stretch}) and (\ref{e:U_bend}) allow for $E$, $\nu$ and the thickness $h$ to vary over the surface. We shall here only consider variations in thickness that respect the axis of symmetry of the shell, {\em i.e.} $h=h(\theta)$. For simplicity, a two--parameter quadratic form for $h(\theta)$ is employed,

\begin{equation}
h(\theta)= A(\h,\rho)+B(\h,\rho)\theta^{2}\quad.
\label{e:h_theta}
\end{equation}

\noindent{}Here, the constants $A$ and $B$ are completely specified by the mean thickness $\h$ and a dimensionless non-uniformity parameter $\rho=[h(\tn)-h(0)]/h(\tn)$. In this way, $\rho=0$ corresponds to uniform shells, and $\rho>0$ to shells that are thinner near the apex than the contact line. The apex thickness vanishes in the limit $\rho\rightarrow 1^{-}$.

Equilibrium configurations ${\mathcal S}$ for given shell parameters and pressure were found by numerically minimising the total energy using the BFGS variable metric method~\cite{NumericalRecipes}. This was performed on the displacement vector fields $u_{r}(\theta)$ (radial displacement) and $u_{\theta}(\theta)$ (angular displacement), where the $\theta$ range was discretised into 50---100 segments as the situation dictated. The midplane strain tensor $u_{ij}$ was found by evaluating the full, (geometrically) non--linear expression $2u_{ij}=\partial_{i}u_{j}+\partial_{j}u_{i}+(\partial_{i}u_{k})(\partial_{j}u_{k})$ in spherical coordinates~\cite{Landau&Lifshitz}. For simplicity, only the linear expressions for $\kappa_{ij}$~\cite{SeideBook} were used, since intuitively it is non--linearity in the in--plane strains that controls buckling. The severity of this simplification was checked by comparing simulation data to known results for spherical caps ({\em i.e.} the buckling pressure $P_{\rm c}$ for uniform $h$~\cite{KaiYuan:1994b,Grigolyuk:2003}), for which the agreement was good.

The results presented below are for a Poisson ratio $\nu=0.3$ and clamped boundary conditions, in which the angle the shell makes with the contact line is fixed. For robustness, we have also simulated a representative sample of systems with $\nu=0.5$ (incompressible), $\nu=0$ and $\nu=-0.5$, and found no qualitative change in the results, merely a modulation of the master curves (discussed below) amounting to no more than 10\%, although inspection of the shell equations (\ref{e:U_stretch}-\ref{e:U_bend}) suggests values of $\nu$ close to $-1$ might produce more significant deviations. For hinged ({\em i.e.} no angle constraint) boundary conditions the effect was larger, reducing 
$\lambda^{\rm cusp}$ (again, see below) by $\approx40\%$, and similar sized changes in the critical pressure. Furthermore the master curves for non--uniform shells were primarily shifted in pressure, only weakly in~$\lambda$. Considerations of the microscopic nature of the crust--substrate interaction suggest clamped boundaries to be more realistic, and so we focus our attention on these.

%
% RESULTS
%
\section{Results}
\label{s:results}

\begin{figure}
\includegraphics[width=8.5cm]{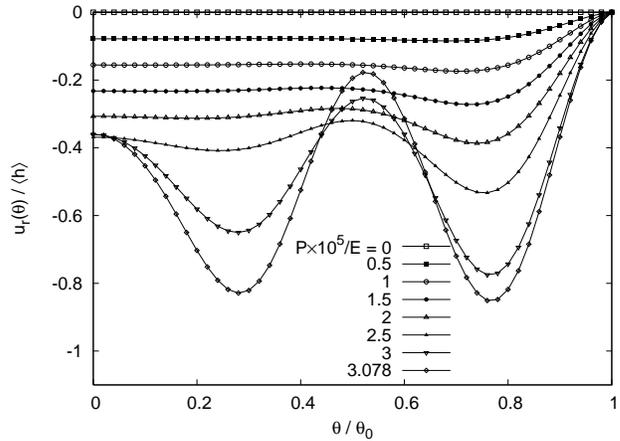}
\caption{
Example of the evolution of the shell deformation with pressure. Each curve corresponds to the outward radial displacement $u_{r}(\theta)$, scaled by the uniform shell thickness $\h=h=5\times10^{-3}R$, for different pressures~$P$. The final value of $P$ is just prior to buckling. In this example, $\theta_{0}=0.5$, $\nu=0.3$ and $E$ is the Young's modulus of the shell material.
}
\label{f:mode_evolve}
\end{figure}

For small pressures~$P$, the shell displacements are linear in $P$ and the elastic strain energy is simply proportional to~$P^{2}$. Furthermore the shape of the shell always takes the lowest mode consistent with the boundary conditions, namely the $n=1$ `dimple' form. Higher modes only arise when the non--linear terms in the membrane strain become important. This is demonstrated in Fig.~\ref{f:mode_evolve}, which shows the radial displacement as a function of $\theta$ for different pressures $P$ from $0$ up to the buckling point. Defining the mode $n$ to be the number of turning points in the radial displacement $u_{r}(\theta)$ in the range $0\leq\theta<\theta_{0}$, it is evident that $n=1$ for small $P$ but takes a higher value just before buckling, $n=4$ in this example. Note that $n$ does {\em not} pass through any intermediate states $n=2$ or $n=3$; the mode $n$ of the shell deformation is fixed once non--linear effects become established, only the amplitudes continue to vary with~$P$. This is typical of the behaviour observed in the parameter range studied. Thus when below we present results for $n$ just prior to buckling, it should be understood that this same mode exists for all prior $P$ down to the linear regime.

The physical mechanism behind the snap--buckling is as follows. In the limit $h\rightarrow0$, the bending energy $\ub$~(\ref{e:U_bend}) becomes arbitrarily small compared to the stretching energy $\us$~(\ref{e:U_stretch}) and so can be ignored (unless the shell deforms inextensibly, which is not possible for this geometry~\cite{SeideBook,Libai&Simmonds}). Now observe that if the deformed shell takes the shape of an {\em inverted} spherical surface with the same radius of curvature~$R$, the midplane strains $u_{ij}$, and hence $\us$, vanish. In between this inverted shape and the undeformed non-inverted shape lay flatter configurations with a high $\us$, which collectively play the role of an energy barrier between the two preferred states. As $P$ is varied, this energy landscape `tilts' until the system hops between the two local minima, {\em i.e.} it {\em snap--buckles}.

In contrast to the stretching energy, the bending energy $\ub$ increases monotonically, even when the shell becomes inverted. Hence as $\h$ increases and $\ub$ becomes more significant, the depth of the energy well corresponding to the inverted shape becomes smaller, until at some point it vanishes. When this happens, there is no buckling and the shell continuously and smoothly deforms for slowly varying~$P$ ({\em i.e.} the $P-\Delta V$ curve is monotonic). Thus it should be possible to identify some dimensionless combination of parameters, including $\h$, that determines whether or not the shell buckles; for $\h$ too large, there should be no instability. This insight is confirmed the simulation results, as we now describe.

\subsection{Critical pressure $P_{\rm c}$ and mode $n$}

\begin{figure}
\includegraphics[width=8.5cm]{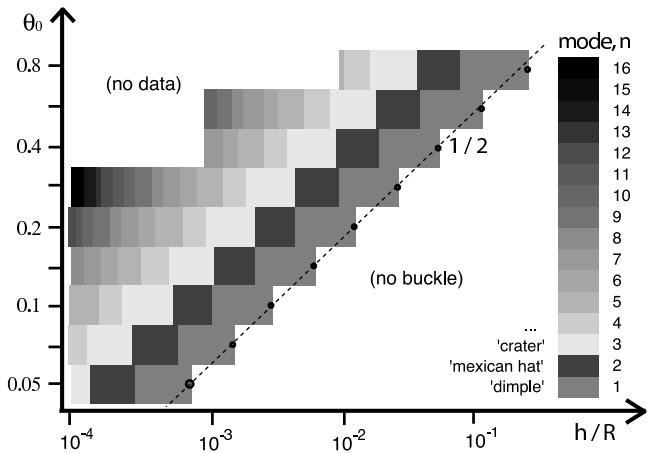}
\caption{
Plot of shell shape just prior to buckling (as given in the key) as a function of contact angle $\theta_{0}$ and shell thickness normalised to the radius of curvature, $h/R$, for a uniform $h(\theta)=\h$. The `blockiness' is due to a large data interval for $\theta_{0}$ of $\approx41\%$. The solid circles denote the thickness at which buckling is first observed, and the dashed line has the slope $1/2$ as shown.}
\label{f:buckle_mode}
\end{figure}

The variation of the mode $n$ prior to buckling with both contact angle $\theta_{0}$ and uniform shell thickness $h/R$ is given in Fig.~\ref{f:buckle_mode}. As expected, for each $\theta_{0}$ there is a critical thickness that separates shells that buckle (thin shells) from those that don't (thick shells). For small angles this line is well approximated by $\h/R\propto\theta_{0}^{2}$, which can be attributed to a crossover from a stretch--energy dominated regime to a bending--energy dominate one; see for instance the scaling theory of~\cite{Landau&Lifshitz} or section~\ref{s:scaling}. Furthermore, the mode $n$ increases the further one moves into the buckling regime, and again the crossover between modes seem to lie on lines $\h/R\propto\theta_{0}^{2}$, an observation that is quantified below. Thus dimpling, mexican hats and so on are just the beginnings of a series of modes that increases apparently without bound.

In the so--called shallow shell regime~$\theta_{0}\ll1$, it is customary to describe the shell geometry by the dimensionless quantity~\cite{ShilkrutBook}

\begin{equation}
\lambda
=
\left[12(1-\nu^2)\right]^{\frac{1}{4}}
\sqrt{\frac{R\tn^{2}}{\h}}
\quad.
\label{e:lambda}
\end{equation}

%known as the `rise height ratio' (since for shallow shells the initial apex height is $R(1-\cos\tn)\approx\frac{1}{2}R\tn^{2}$)
\noindent{}Intuitively, small $\lambda$ corresponds to thick and/or flat shells, which behave something like flat plates, and large $\lambda$ to thin and/or steep shells which deform as curved shells. In addition, the pressure is normalised to the buckling pressure for a full sphere~\cite{Libai&Simmonds},

\begin{equation}
q
=
\frac{P}{P_{\rm c}^{\rm sphere}}
=
\frac{P}{2Eh^{2}/[R^{2}\sqrt{3(1-\nu^{2}})]}
\quad.
\label{e:p}
\end{equation}

\begin{figure}
\includegraphics[width=8.5cm]{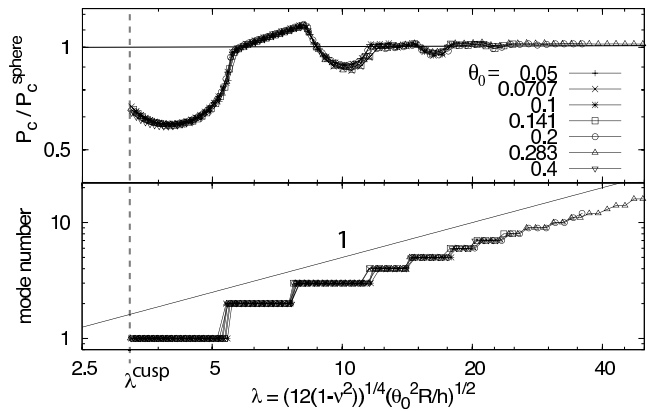}
\caption{
Buckling pressure $P_{\rm c}$ scaled to the full sphere solution $P_{\rm c}^{\rm sphere}$ {\em (upper panel)} and the mode just prior to buckling {\em (lower panel)} versus $\lambda$ for shells of uniform thickness $h$ and contact angles $\theta_{0}$ given in the key. The data range for each $\theta_{0}$ differ, but {\em all} extend down to $\lambda^{\rm cusp}$.}
\label{f:Pc_n_uniform}
\end{figure}

\noindent{}When plotted in terms of these dimensionless variables, both $P_{\rm c}$ and $n$ collapse onto a single curve, as shown in Fig.~\ref{f:Pc_n_uniform}. Although the highest $\theta_{0}$ were omitted from this plot, deviations only amount to around $\approx15\%$ for $\theta_{0}=0.8$~rads~$\approx45^{\circ}$, suggesting the so--called `shallow shell' limit provides a decent approximation for even somewhat steep contact angles.

It is apparent from this figure that $q_{\rm c}\approx1$ for $\lambda\gg1$, {\em i.e.} the buckling pressure approaches the full sphere solution for large values $\lambda$. Furthermore, the mode $n\sim\lambda$ to good approximation, which is also confirmed by the scaling argument given later. Plotting both sets of data on the same axes as here also shows a correlation between the features of the curves, namely that the `kinks' in the pressure data correspond to increases in the mode. Thus deviations from the complete sphere solution depend on the mode of deformation. As a final observation, note that a minimum $\lambda^{\rm cusp}\approx3.2$ for buckling to occur was observed for each data curve (the label `cusp' is explained below). This, and the shape of the $P_{\rm c}(\lambda)$ curve, lie within 5\% of known results from more rigourous shell simulations~\cite{ShilkrutBook}, justifying the simplifying choice of linear bending expressions in the simulations.

\begin{figure}
\includegraphics[width=8.5cm]{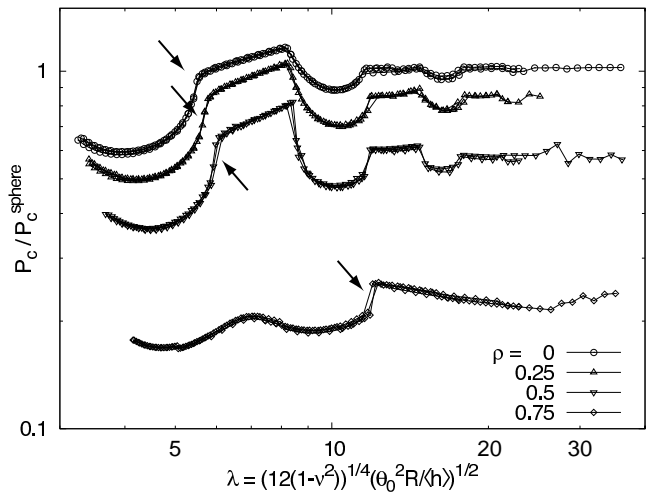}
\caption{
Buckling pressure $P_{\rm c}$ scaled by the full sphere solution, for different values of the non--uniformity parameter $\rho$ given in the key. For each value of $\rho$, data for $\theta_{0}=0.1$, 0.2 and 0.4 has been plotted. To guide the eye, an arrow have been added to each line to indicate the kink corresponding to the $n=1\leftrightarrow n=2$ mode transition.}
\label{f:Pc_non_uniform}
\end{figure}

Introducing a variable shell thickness through the non--uniformity parameter $\rho>0$ does not alter the data collapse described above, but rather shifts and distorts the master curve. $P_{\rm c}$ versus $\lambda$ for different values of $\rho$ is given in Fig.~\ref{f:Pc_non_uniform}, in terms of the same dimensionless quantities as before. There is again good data collapse, but now onto distinct curves, with one such curve for each~$\rho$. As $\rho$ increases, the curves move to lower pressures; in words, shells that are thinner near the apex than the contact line are weaker than uniform shells with the same mean thickness. Furthermore they move to higher $\lambda$, so that $\lambda^{\rm cusp}$ increases and a broader range of shells will not buckle ({\em i.e.} have $\lambda<\lambda^{\rm cusp}$). Note that although the curves for different $\rho$ share the same essential features, including kinks when the mode number changes, they cannot be collapsed onto a single `super--master' curve by simple scaling. Note also that the relationship $n\sim\lambda$ still holds (data not shown).

\begin{figure}
\includegraphics[width=8.5cm]{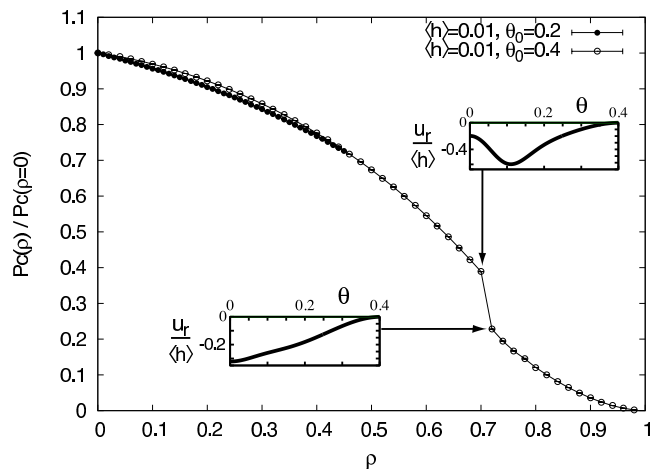}
\caption{
$P_{\rm c}$ as a function of the non--uniformity parameter $\rho$, scaled to the uniform thickness $\rho=0$ case, for mean thickness and contact angle given in the key. The $\theta_{0}=0.2$ shells cease to buckle for $\rho>0.45$. Also, the radial displacement $u_{r}(\theta)$ just prior to buckling for the two shells either side of the `kink' in the $\theta_{0}=0.4$ line are given, showing the decrease in mode.}
\label{f:var_rho}
\end{figure}

An alternative way to probe non--uniform shell thickness is to hold $\lambda$ fixed and instead vary $\rho$. Since the master curves tend to drift to lower $q$ and higher $\lambda$ with $\rho$, increasing $\rho$ should act to {\em decrease} the mode number (possibly removing buckling) and lower the critical pressure. This is precisely what is observed; as evident in Fig.~\ref{f:var_rho},  $P_{\rm c}$ is a decreasing function of~$\rho$, and furthermore mode reduction or loss of buckling can occur at finite values of $\rho>0$. Thus a non--uniform shell, rather than increasing the mode number (as perhaps might be expected), on the contrary acts to {\em decrease} the mode. Quantifying the dependency on $\rho$ is difficult as it depends on the details of the family of master curves parameterised by $\rho$, but we note that the data for the $\theta_{0}=0.4$ case in the figure appears to scale as $P_{\rm c}(\rho)\sim(1-\rho)^{2}$ for small $1-\rho$. We have no explanation for this at present.

\subsection{The onset of buckling: Cusp scaling}

\begin{figure}
\includegraphics[width=8.5cm]{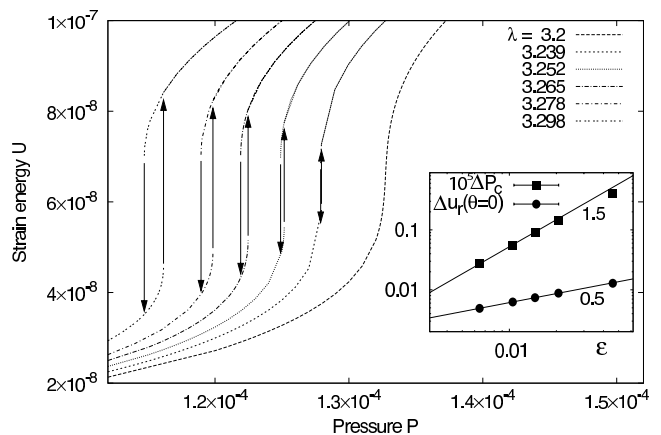}
\caption{
Strain energy versus pressure $P$ for $\theta_{0}=0.2$ and the $\lambda$ given in the key, in units such that $R=E=1$. The vertical arrows denote critical pressures for snap--through buckling under increasing~$P$ (upwards--pointing arrows), and snap--back buckling when $P$ is subsequently decreased (down arrows). {\em (Inset)}~Difference in the two critical pressures $\Delta P_{\rm c}$ and the deflection of the apex just prior to buckling $\Delta u_{r}(0)$ as a function of $\varepsilon=\lambda-\lambda^{\rm cusp}$ with $\lambda^{\rm cusp}\approx3.231$. The solid lines have slopes of 1/2 and 3/2 as shown.}
\label{f:cusp}
\end{figure}

Although evaporation is clearly a unidirectional process, it is nonetheless instructive to consider cyclic variations around the buckling point since, as evident from Fig.~\ref{f:schematic}, such a protocol probes the hysteresis curve and allows greater insight to be gained into the nature of buckling singularity. A sample of hysteresis curves for a range of $\lambda$ near $\lambda^{\rm cusp}$ are given in Fig.~\ref{f:cusp}. The `width' of the hysteresis loop was taken to be the difference between the two critical pressures for snap--through and snap--back buckling, $\Delta P_{\rm c}$. For the `height', either the difference in strain energy $\Delta{\mathcal U}$, or the difference in apex displacement $\Delta u_{r}(\theta=0)$, between pre--buckled and post--buckled states can be used; both quantities give the same scaling. As evident from the figure, the hysteresis loop vanishes continuously as $\lambda\rightarrow\lambda^{\rm cusp}$, according to $\Delta P_{\rm c}\sim(\lambda-\lambda^{\rm cusp})^{3/2}$ and $\Delta u_{r}\sim(\lambda-\lambda^{\rm cusp})^{1/2}$.

These exponents are characteristic of a {\em cusp} singularity~\cite{Thompson&Hunt,GilmoreBook}, and indeed the scaling theory of section~\ref{s:scaling}, which explicitly has a potential energy function capable of generating a cusp, also predicts the same exponents (see below for details). This explains our use of the superscript {\em `cusp'} when referring to this singularity. The picture is as follows: for $\lambda>\lambda^{\rm cusp}$, the system encounters two buckling transitions on suitably varying $P$, one corresponding to forward ({\em `snap--through'}) buckling, the other to reverse or {\em `snap--back'} buckling. As $\lambda^{\rm cusp}$ is approached, these two transitions approach until `annihilating' at a higher--order singularity, namely the cusp point. For $\lambda<\lambda^{\rm cusp}$ there is no buckling and the shell deforms continuously for all pressures.

%
% THEORY
%
\section{Scaling theory}
\label{s:scaling}

It is possible to construct a scaling argument to describe and compare to the simulations results. This is similar in style to that in Landau and Lifshitz~\cite{Landau&Lifshitz}, but with the addition of explicit non--linear terms for the stretching energy, and a new variable to mimic the mode. We shall also present some simple, linear calculations for non-uniform shells.

\subsection{Variation with mean thickness: Buckling}

The approximations employed in setting up the scaling theory are drastic but far--reaching. Firstly, only the radial component of the displacement $u_{r}$ is considered; the angular component $u_{\theta}$, which corresponds to in-shell motion and is anyway small, is simply neglected. Secondly, the angular dependence $u_{r}(\theta)$ is subsumed into a characteristic {\em inward} displacement $\zeta\sim - u_{r}$, with an unknown (but positive as defined here) dimensionless prefactor. We further define a dimensionless variable $\omega$ through the characteristic derivatives $\partial u_{r}/\partial\theta\sim - \zeta\omega/\tn$ and $\partial^{2}u_{r}/\partial\theta^{2}\sim - \zeta(\omega/\tn)^{2}$. We shall below identify $\omega$ with the mode number~$n$, although $\omega$ is a continuous variable and as such potentially contains more information than the discrete~$n$. %We also assume that the amplitude of the modulations in $u_{r}(\theta)$ are much smaller than the amplitude of the characteristic $\zeta$.

In terms of the dimensionless quantities $\lambda$ already defined in~(\ref{e:lambda}), $p=P/(E\theta_{0}^{4})$ and $x=\zeta/\h$, the characteristic energies per unit surface area (including the work $W$ done by the pressure) can be derived from (\ref{e:U_stretch}-\ref{e:U_bend}),

\begin{eqnarray}
&\delta\us&=E\h\tn^{4}
\left\{C^{\rm s}_{2}\frac{x^{2}}{\lambda^4}-C^{\rm s}_{3}\frac{\omega^{2}x^{3}}{\lambda^{6}}
+C^{\rm s}_{4}\frac{\omega^{4}x^{4}}{\lambda^{8}}
\right\}
\nonumber\\
&\delta\ub&=\frac{E\h\tn^{4}x^{2}}{\lambda^{8}}
\left\{
%f(\omega)
C^{b}_{2}\omega^{2}+\nu C^{b}_{3}\omega^{3}+C^{b}_{4}\omega^{4}
\right\}
\nonumber\\
&\delta W&=-C_{w}pE\tn^{4}\h x
\label{e:scaling}
\end{eqnarray}

\noindent{}where all of the $C_{i}$'s are unknown, but positive, dimensionless constants. As in the simulations, the non--linear strain $u_{ij}$ but linear curvatures $\kappa_{ij}$ have been used. These equations must be supplemented by the constraint that $\omega$ is bounded below by some positive ${\mathcal O}(1)$ constant, which is required for consistency with the fixed boundary conditions. The equilibrium $x$ and $\omega$ for a given $p$ can be found by minimising the total energy.

Consider first the linear response $x\ll1$, for which terms in $x^{3}$ and higher can be ignored and the energy function is quadratic in~$x$. Then (\ref{e:scaling}) predicts a negative $\omega$, which is not consistent with the boundary conditions, so we fix $\omega$ to be an ${\mathcal O}(1)$ constant and minimise for $x$ alone. In the bending--dominated regime $\lambda\ll1$, when $\delta\us$ can be neglected, $x\sim p\lambda^{8}$ or $\zeta\sim P\tn^{4}R^{4}/E\h^{3}$. Conversely, in the stretching--dominated regime $\lambda\gg1$, $x\sim p\lambda^{4}$ or $\zeta\sim PR^{2}/E\h$. The crossover between these two regimes lies at $\lambda={\mathcal O}(1)$, {\em i.e.} $\theta_{0}\sim\h/R$, 
which coincides with the onset of buckling observed in the simulations. This confirms the intuitive picture that buckling arises when the stretching term dominates.

Let us now turn to non--linear effects and buckling. For $x={\mathcal O}(1)$, the energy (\ref{e:scaling}) is quartic and thus admits 1 or 3 equilibrium solutions. This is most clearly seen in the stretching regime $\lambda\gg1$, when the strain energy is dominated by $\delta\us$ and all 3 roots exists, two minima at $x=0$ and $x\sim\lambda^{2}/\omega^{2}$ and an unstable local maximum inbetween. As $p$ increases, the root initially at $x=0$ moves to higher $x$ until it and the local maximum annihilate each other; this is the critical pressure when the system jumps to the second minimum. It is straightforward to show that, just before this buckling event, $x_{\rm c}={\mathcal O}(1)$, $\omega_{\rm c}\sim\lambda$ and $p_{\rm c}\sim\lambda^{-4}$, or in non-normalised variables, $\zeta_{\rm c}\sim\h$, $\omega_{\rm c}\sim\theta_{0}\sqrt{R/\h}$ and $P_{\rm c}\sim E\h^{2}/R^{2}$. This is in agreement with the simulation results discussed above if we regard $\omega$ as a continuous analogue of the mode number~$n$. It should be mentioned here that the second root and local maximum do not exist for {\em all} values of the $C^{\rm s}_{i}$. Since it is beyond this scaling theory to determine the $C^{\rm s}_{i}$, we must simply assume they take values such that 3 roots exists for $\lambda\gg1$.

For $\lambda={\mathcal O}(1)$, the size of the jump to the second minimum at $p_{\rm c}$ is smaller than for large $\lambda$, and vanishes at a finite value $\lambda^{\rm cusp}$. The corresponding critical pressure, $p^{\rm cusp}_{\rm c}$, and $\lambda^{\rm cusp}$ describe a point in parameter space for which all three equilibria (including the unstable maximum) coincide. To analyse scaling behaviour around this point, we define $\varepsilon=\lambda-\lambda^{\rm cusp}$ and again assume that $\omega$ is a fixed, ${\mathcal O}(1)$ constant (this is consistent with the simulations, for which the lowest mode solution was always observed around the critical $\lambda$). By expanding the energy around the cusp point in $\varepsilon$ and preserving only the lowest order terms, it is straightforward to derive the scaling observed in the simulations, namely that the difference between the two critical pressure (for snap--through and snap--back buckling) scales as $\Delta p^{\pm}_{\rm c}\sim\varepsilon^{3/2}$, and the corresponding configurations scale as $\Delta x^{\pm}_{\rm c}\sim\varepsilon^{1/2}$. These exponents and the energy function (\ref{e:scaling}) describe a cusp catastrophe~\cite{GilmoreBook,Thompson&Hunt}, as suggested by our notation.

\subsection{Non--uniform thickness}

The nature of scaling theory means that it is not well suited to describing spatially--varying quantities. Nonetheless a simple argument, in which bending energy and non--linearities are ignored, can be treated analytically, as we now describe. For $\rho>0$, the apex is thinner than the `wing' region near the contact line, and might be expected to deform more. We therefore define two characteristic deformations, $\zeta_{\rm a}$ and $\zeta_{\rm w}$, for the apex and wing regions respectively. These two regimes are considered to be sharply separated at an angle $\theta_{1}<\theta_{0}$, which defines the third variable of this model. For any given pressure~$P$, the values of $\zeta_{\rm a}$, $\zeta_{\rm w}$ and $\theta_{1}$ are found by minimising the linear strain energy restricted to stretching terms,

\begin{eqnarray}
{\mathcal U}&=&
C_{\rm a}E\zeta^{2}_{\rm a}\int_{0}^{\theta_{1}}{\rm d}\theta\,h(\theta)\sin\theta
+
C_{\rm b}E\zeta^{2}_{\rm w}\int_{\theta_{1}}^{\theta_{0}}{\rm d}\theta\,h(\theta)\sin\theta
\nonumber\\
&&
-C_{\rm ap}P\zeta_{\rm a}R^{2}\int_{0}^{\theta_{1}}{\rm d}\theta\,\sin\theta
-C_{\rm wp}P\zeta_{\rm w}R^{2}\int_{\theta_{1}}^{\theta_{0}}{\rm d}\theta\,\sin\theta
\nonumber\\
\label{e:two_mode}
\end{eqnarray}

\noindent{}where as before the $C_{i}$'s are ${\mathcal O}(1)$ dimensionless constants. Substituting the explicit quadratic form for $h(\theta)$~(\ref{e:h_theta}) is most easily treated by considered small $\rho$ and $\rho\approx1$ separately, and again we assume shallow shells $\theta_{0}\ll1$.

For $\rho\ll1$, {\em i.e.} almost--uniform shells, $\theta_{1}/\theta_{0}\approx(1-\rho/8)/\sqrt{2}$ and that the deformation around the apex is enhanced with respect to the wing by a small amount $\zeta_{\rm a}/\zeta_{\rm w}\approx 1+\rho/2$, to leading order in~$\rho$. The critical pressure, which  can be estimated in linear theory as when the shell becomes flat, is similarly reduced, $P_{\rm c}(\rho)/P_{\rm c}(0)\approx(1-\rho/4)$. Although this correctly predicts an initially linear decrease in $P_{\rm c}$ with $\rho$, it is clear from the simulations that the slope depends on the shell parameters rather than taking the fixed value $-1/4$ as suggested here.

Repeating the procedure for $\rho\approx1$ and expanding in $1-\rho\ll1$ reveals a more drastic modification from the uniform case. Now $\theta_{1}/\theta_{0}\approx(1-\rho)^{1/4}$ and $\zeta_{\rm a}/\zeta_{\rm w}\approx(1-\rho)^{-1/2}$, so a small region around the apex becomes strongly deformed. The small exponent of $1/4$, however, suggests $\rho$ must be very close to $1$ for this regime to be clearly observed. The critical pressure vanishes as $P_{\rm c}(\rho)/P_{\rm c}(0)\approx(1-\rho)^{a}$ with $a=1$, whereas the simulations instead suggest $a=2$ as already described, again demonstrating this argument is at best capable of producing qualitatively correct predictions only. It is, however, not clear at this time how to improve the theory with regards non--uniform shells while retaining the appealing clarity of the scaling approach.

%
% DISCUSSION
%
\section{Discussion}
\label{s:discussion}

The findings outlined in this article demonstrate that the effects of non--uniform shell thickness on the deformation and buckling of elastic caps can be handled in a systematic manner, once the mean shell thickness $\h$ has been factored out of the thickness profile~$h(\theta)$. For the quadratic profile considered here, the dimensionless non--uniformity parameter $\rho$ was shown to determine the behaviour of the system, in that each $\rho$ corresponded to a different master curve once the various quantities have been normalised in a suitable manner. Based on this finding, we hypothesise that {\em any} non--uniform thickness profile $h(\theta)=\h f(\theta;\theta_{0})$ will produce similar behaviour, with master curves parameterised by $f(\theta;\theta_{0})$. The overall form of the curve, such as kinks corresponding to changes in the mode, is also expected to remain the same, with only quantitative details differing.

The issue then becomes one of determining the actual thickness profile generated by droplet evaporation of sessile polymer or colloid solutions. This is far from trivial; crust formation involves a number of physical processes~\cite{deGennes:2002}, including vapour diffusion, solvent flow and gelation, any and all of which may overlap in time with the elastic deformation considered in isolation here. Until such a time that these effects are properly quantified for the droplet geometry, it is difficult to provide any definite predictions. Nonetheless some broad statements can be made based on our findings. For instance, we expect that shells that are thinner near the apex than the base will generally have a lower mode number $n$ that uniform shells of the same mean thickness. This, perhaps ironically, suggests that, if a uniform deposit profile is required, it may be advantageous to have a {\em non--uniform} thickness, since lower mode numbers $n$ correspond to flatter final shapes. Also such shells are {\em weaker}, in that the critical pressure is lower, suggesting they will reach non--linear region of the $P$-$\Delta V$ curve comparatively quickly. %Although they will not buckle here if under controlled volume, this still helps to identify the regime where non--linear effects have become fully established and the mode $n$ prior to buckling will arise.

Notwithstanding the difficulties associated with the full evaporating droplet problem, it may be worthwhile to briefly speculate here on the effects of ongoing crust formation during deformation. Supposing that polymer is added to the crust in an initially unstressed state, the effects of continued gelation will be to reduce the mean shell stress, and alter the intrinsic curvature towards the current value. Given that the selection of higher modes requires non--linear deformations from some (unstressed) reference configuration, we speculate that the mode number $n$ will be {\em reduced} by this effect. However, this assumes the increase in crust thickness to be uniform. In reality, it is expected that the rate of deposition of new material to the shell will be higher for regions that are convex relative to the solution (that is, that `bulge' into it) than for concave regions, since convex regions will be exposed to a greater volume of solution and thus solute. This suggests a two--way coupling between deformation and crust thickening, a potentially rich problem for which further study would be desirable.

When comparing our work to the experiments of Pauchard {\em et al.}~\cite{Pauchard:2003a,Pauchard:2003b,Pauchard:2003c}, Gorand {\em et al.}~\cite{Gorand:2004} and Kajiya {\em at al.}~\cite{Kajiya:pre}, a difference in notation should be observed: in these works the term `instability' is used to refer to the formation of any mode $n>1$, whereas we reserve the term for snap--buckling events (also symmetry--breaking deformations). From our findings, modes with $n>1$ (such as mexican hats) form when the characteristic deformation is of the order of the thickness of the shell, and thus will be delayed with respect to the formation of the crust itself. Nonetheless the sequence of deformed shapes with increasing $\theta_{0}$ observed in~\cite{Gorand:2004} appears to be consistent with our work, although without any theory for the formation of the crust, it is difficult to provide any close correspondence with experiments. For instance, no $n>3$ mode has been observed in experiments; this may be due to the emergence of asymmetric modes (which for uniform shells are known to occur by $\lambda\approx8$~\cite{ShilkrutBook}, {\em before} the emergence of even the $n=3$ `crater' mode), due to dynamic effects, or more simply that the required shell parameters are outside of the experimental parameter window.

\section*{Acknowledgements}

The author would like to thank M. Doi, T. Yamaue and S. Komura for useful discussions. This work was funded by the JSPS award no. P04727.

%
% REFERENCES - use standard "\cite{}" command
%

\end{document}